\documentclass[conference]{IEEEtran}

\usepackage{graphicx} %%For loading graphic files
\usepackage{amsmath}
\usepackage{amsthm}
\usepackage{amsfonts}
\usepackage{multirow}
\usepackage{rotating}
\usepackage{setspace}
\usepackage{array}

\usepackage[font=footnotesize]{subfig}
\begin{document}
\pagestyle{empty} %No headings for the first pages.
\title{Autonomic Management of Maintenance Scheduling in Chord}

\author{\IEEEauthorblockN{Markus Tauber, Graham Kirby and Alan Dearle}
\IEEEauthorblockA{School of Computer Science\\
University of St Andrews\\
Fife, Scotland KY16 9SX\\
Email: \{markus, graham, al\}@cs.st-andrews.ac.uk}
}

\maketitle

\begin{abstract}
This paper experimentally evaluates the effects of applying autonomic management
to the scheduling of maintenance operations in a deployed Chord network, for
various membership churn and workload patterns. Two versions of an autonomic
management policy were compared with a static configuration. The autonomic
policies varied with respect to the aggressiveness with which they responded to
peer access error rates and to wasted maintenance operations. In most
experiments, significant improvements due to autonomic management were observed
in the performance of routing operations and the quantity of data transmitted
between network members. Of the autonomic policies, the more aggressive version
gave slightly better results.
\end{abstract}

\section{\label{intro}Introduction}

Various peer-to-peer (P2P) overlay networks, such as Tapestry \cite{tapestry},
CAN \cite{CAN}, Pastry \cite{pastry} and Chord \cite{chord}, support the
key-based routing (\emph{KBR}) abstraction \cite{commonapi}. This allows any
given key value to be mapped to a live node in the network, in the presence of
dynamic change in the network membership. Each node maintains knowledge of the
addresses of some subset of network nodes, its \emph{peer-set}. These links
may be categorized as follows:

\begin{enumerate}
\item those that are needed for correct routing
\item those that are useful for efficient routing
\item those that may be needed to repair network topology
\end{enumerate}

For example, a node's peer-set in Chord includes its successor (category 1 and
category 3), predecessor (category 3), successor list (category 3) and fingers
(category 2).

Links in categories 1 and 2 are used during routing to locate the target node for
a given key in a series of hops. As network membership changes, various links may
come to refer to failed nodes, or they may become incorrect with respect to
correct routing (category 1), efficient routing (category 2), or possible later
overlay repair needs (category 3).

To maintain routing correctness and efficiency in the presence of membership
churn, errors must be detected and rectified. A node may discover an error during
a routing operation, be notified of it by another node, or discover it through a
periodic checking process. Once an error is discovered, a new target is
established and the referring link updated.

Here we focus on the scheduling of periodic checking processes, that is,
controlling the rate at which checks are made. Each check requires one or more
(attempted) inter-node interactions. The optimal scheduling of such maintenance
operations depends on dynamic factors, including the workload---the pattern of
routing calls applied to the network---and the churn in network membership.
Optimality of scheduling may be considered with respect to various properties of
the overlay network, including user-perceived routing performance, and resource
consumption.

Resource consumption can be optimized in isolation by setting the frequency of
maintenance operations to zero. With respect to performance, the ideal rate
depends on the current workload and churn rate. If the maintenance rate is too
high then performance suffers due to wasted CPU and network resource; if it is
too low then performance is also reduced, due to higher routing cost. This
increase in routing cost arises due to communication errors resulting from
attempts to follow broken links, leading to retries, and due to the use of
functioning but non-optimal links.

The maintenance rate giving the best trade-off between resource consumption and
performance depends on the prevailing conditions. For example, if no routing
calls are made, or the network membership is completely static, then the optimal
behavior is to perform no maintenance, since it represents pure overhead.
Conversely, under a heavy or varying workload, or rapid network churn, it may be
beneficial for nodes to expend significant maintenance effort in order to sustain
high performance for routing operations.

In most P2P protocols, maintenance operations are scheduled at a statically
configured fixed rate. Even when the workload and churn remain relatively
constant throughout the lifetime of a network, a statically configured rate is
unlikely to be optimal for that particular combination of workload and churn.
Furthermore, workload and churn may vary dynamically. Even if the statically
configured rate happens to be appropriate for the initial circumstances, it may
become less suitable as conditions vary.

We investigated the use of autonomic management \cite{vision_AC_kephart} to
control maintenance scheduling in response to dynamically changing conditions. We
hypothesized that under non-changing conditions this would allow the system to
converge on a configuration that was more suitable than any that could be set
\emph{a priori}. Furthermore, the system would be able to react to changes in
conditions by dynamically adopting more appropriate configurations.

We designed a range of scheduling management policies for Chord, and evaluated
them on a small experimental test-bed, in comparison to a fixed maintenance
schedule. 
We decided to focus on network usage as the resource consumption measure most
likely to be significantly affected.
The effects on elapsed routing time and bandwidth consumption between
peers were measured for various workloads and churn patterns.

The best autonomic management policy led to an improvement in both performance
and resource consumption for 75\% of the workload/churn combinations tested. For
25\% of combinations the policy led to an improvement in one metric and worsening
in the other; in none of the combinations did it lead to worsening in both
metrics.

The paper is structured as follows: related work is discussed in section
\ref{back_a_rel}. The details of our autonomic manager for Chord are given in
section \ref{autonomic_management}. Section \ref{validation} outlines the design
of the experiments, while the experimental results are presented in section
\ref{results}.

\section{\label{back_a_rel}Related Work}

\subsection{Dynamic Adaptation of Maintenance Scheduling}

The most closely related work is based on Pastry
\cite{controlling_the_cost_of_reliability} and describes the optimization of
resource consumption for a given acceptable message loss rate. This is a form of
performance metric.
Each node dynamically estimates the overall node failure
rate and the size of the overlay, and uses an analytical model to deduce an
appropriate maintenance rate that should yield the target loss
rate. Our approach is simpler in that it does not require estimation of any
global properties, instead applying simple heuristics based on local
observations.

In contrast to our approach, \cite{controlling_the_cost_of_reliability} does not
take user workload into account. Furthermore, it imposes a lower bound on
resource consumption, since the maintenance rate is not further reduced once the
acceptable loss rate is achieved, even in situations where resource consumption
could be lowered while still meeting the target loss rate.

\subsection{Other Approaches to Optimizing Performance and Resource Consumption}

Binzenh{\"o}fer and Leibnitz \cite{estimating_churn_binzenhoefer} describe how
churn may be estimated in Chord networks in order to set maintenance rates
appropriately. They do not however perform any adaptation. Churn is estimated by
monitoring changes in a node's peer-set. The focus is on limiting the probability
of network partition before the next maintenance operation. This approach does
not take user workload into account; the churn estimator receives only
information gathered during maintenance operations. Consequently, the algorithm
may be slow to react to a sudden increase in churn occurring during a low-churn
period with low maintenance rates.

\cite{kademlia-binzenhoefer} proposes modifications to the Kademlia
\cite{kademlia} protocol to propagate information about failed peers and peer-set
membership. This information is propagated at a rate related to network churn,
but the maintenance rate is not controlled explicitly.

Chord2 \cite{chord2} aims to reduce maintenance costs by introducing a two-level
structure, with a smaller ring of high performance super-peers used to manage
finger tables. There is no dynamic control of maintenance intervals.

FS-Chord \cite{FS-Chord} reduces maintenance work caused by unstable nodes
joining for a short period, by only allowing a node to become a full member after
some fixed interval of reliable operation. This approach does not, however,
enable Chord to adapt to changes in node behavior after the initial monitoring
period has passed.

In \cite{onTheStabilityOfChordBinzenhoefer2004} a Chord network model is
developed, showing that increasing the size of the successor list improves
stability. This focuses on network resilience rather than performance. The paper
suggests dynamic adaptation of the successor list length, but the mechanism is
not further specified or evaluated.

\cite{autonommicallyImprovingSecurityAndRobustnessOfStructuredP2POverlaysBinzenhoefer2006}
suggests ``an adaptive mechanism that increases the stabilization
period if the number of known successors shrinks or if the overlay structure is
measured to be more dynamic'' for Chord. Presumably \emph{decreasing} the
stabilization period is intended. No implementation details or experimental evaluation are
given.

In \cite{ChordStabilizationInHighChurn} a modified \emph{stabilize} algorithm
that can improve stability in a Chord network in the presence of high churn is
described. It is suggested that a high maintenance rate is desirable in networks
with high membership churn, and thus there is a correlation between degree of
churn and optimal maintenance rate. However, the paper does not propose any
dynamic adaptation of the maintenance rate.

In \cite{analysis_of_the_evolution_of_p2p_systems} Chord's maintenance mechanism
is analyzed, concluding that the rate is an important configuration factor. The
authors analyze the correlation between maintenance rate and performance, stress
the importance of conservative network usage, and ask whether an optimum
maintenance rate can be learned.

\section{\label{autonomic_management}Autonomic Management of Maintenance Scheduling}

\subsection{Overview}

We developed an autonomic management mechanism for dynamically controlling
maintenance scheduling in Chord nodes, in response to conditions experienced. An
autonomically managed system consists of a target system (in this case, an
individual Chord node) to which an autonomic manager is attached, in order to
dynamically control specific system parameters (in this case, the maintenance
rate for that node).

Our manager executes an autonomic control loop \cite{vision_AC_kephart} which
involves four phases:

\begin{itemize}
\item A \emph{monitoring} phase, during which the manager receives information
about the target system in the form of \emph{events}.
\item An \emph{analysis} phase, during which events are aggregated and a
representation of the current situation is constructed, in the form of abstract
\emph{metrics}.
\item A \emph{planning} phase, during which decisions are made about how to react
to the current situation, driven by \emph{policies}.
\item An \emph{execution} phase, during which the planned actions are carried
out.
\end{itemize}

Our manager is intended to detect when maintenance effort is being wasted and to
decrease the current maintenance rate accordingly. Conversely, it increases
the rate in situations when more vigorous maintenance is appropriate.

The potentially conflicting goals of reducing effort and increasing performance
are managed by two sub-policies. During the planning phase, each sub-policy makes
its own independent recommendation as to how the current maintenance rate should
be adjusted. The mean value of these recommendations is then applied during the
execution phase.

The rationale for this structure is that although the sub-policies will rarely
agree, their recommendations will cancel out in situations where little action is
required, whereas in more extreme situations one will outweigh the other, due to
the magnitude of the recommendations.

During each planning phase, each sub-policy considers metric values derived from
events received during the current autonomic cycle. These events are based solely
on locally gathered data, thus no additional network traffic is generated by the
autonomic manager.

\subsection{Monitoring}

An event is generated whenever:

\begin{itemize}
\item a maintenance operation is executed without any effect on the peer-set
\item a failed attempt to access a peer-set element is made, either during
routing or during a maintenance operation
\end{itemize}

\subsection{Analysis}

Two metrics aggregate the events. The \emph{Wasted Maintenance Count} ($WMC$) and
\emph{Error Count} ($EC$) metrics count the numbers of each event type during the
current autonomic cycle.

$WMC$ models the amount of effort invested in maintenance operations without
effect. A high value suggests that a lot of network traffic was unnecessary,
since either little churn occurred, or maintenance was executed frequently enough
to compensate for such changes. Conversely, a small value implies that the
network traffic due to maintenance operations was effective in correcting
errors, and may therefore be regarded as justifiable.

$EC$ models the accuracy of the peer-set as perceived by user or maintenance
activities attempting to use it. A high value suggests that a large proportion of
the peer-set is not valid. Conversely, a low value suggests either a high degree
of accuracy in the peer-set, due to frequent maintenance or low churn, or low
user demand due to a light workload.

\subsection{\label{subsec:policies}Planning}

Each of the sub-policies is driven by one of the metrics. The sub-policy
concerned with reducing effort considers $WMC$, while the sub-policy concerned
with improving performance considers $EC$. The rationale for the former is
straightforward; for the latter, the point is that user-level routing operations
will retry when errors are encountered, thus a high error rate leads directly to
poorer performance.

As with any negative feedback approach, each sub-policy recommends a change to
the current maintenance rate, of a magnitude related to the difference between
the current value of the relevant metric and some ideal value for that metric.
The further that the metric diverges from the ideal, the more aggressive the
response that is recommended.

Since both metrics count undesirable events, the ideal value for both metrics is
zero. Whenever $WMC$ is non-zero the sub-policy concerned with reducing effort
recommends a reduction in the maintenance rate, while whenever $EC$ is non-zero
the sub-policy concerned with improving performance recommends an increase in the
maintenance rate.

For ease of integration with Chord, in practice each sub-policy recommends a new
interval between maintenance operations, rather than a rate. It calculates the
proportion $P$ by which the current interval should be changed. The new interval
is then calculated, for the sub-policy using $WMC$, as:

\vspace{-0.2cm}
\begin{equation}
new\ interval = current\ interval \times (1 + P)
\end{equation}

The sub-policy using $EC$ seeks to increase the maintenance rate, so the new
lower interval is calculated as:

\vspace{-0.2cm}
\begin{equation}
new\ interval = current\ interval \times (1 - P)
\end{equation}

\begin{figure}[b]
  \centering
    \includegraphics[width=8cm]{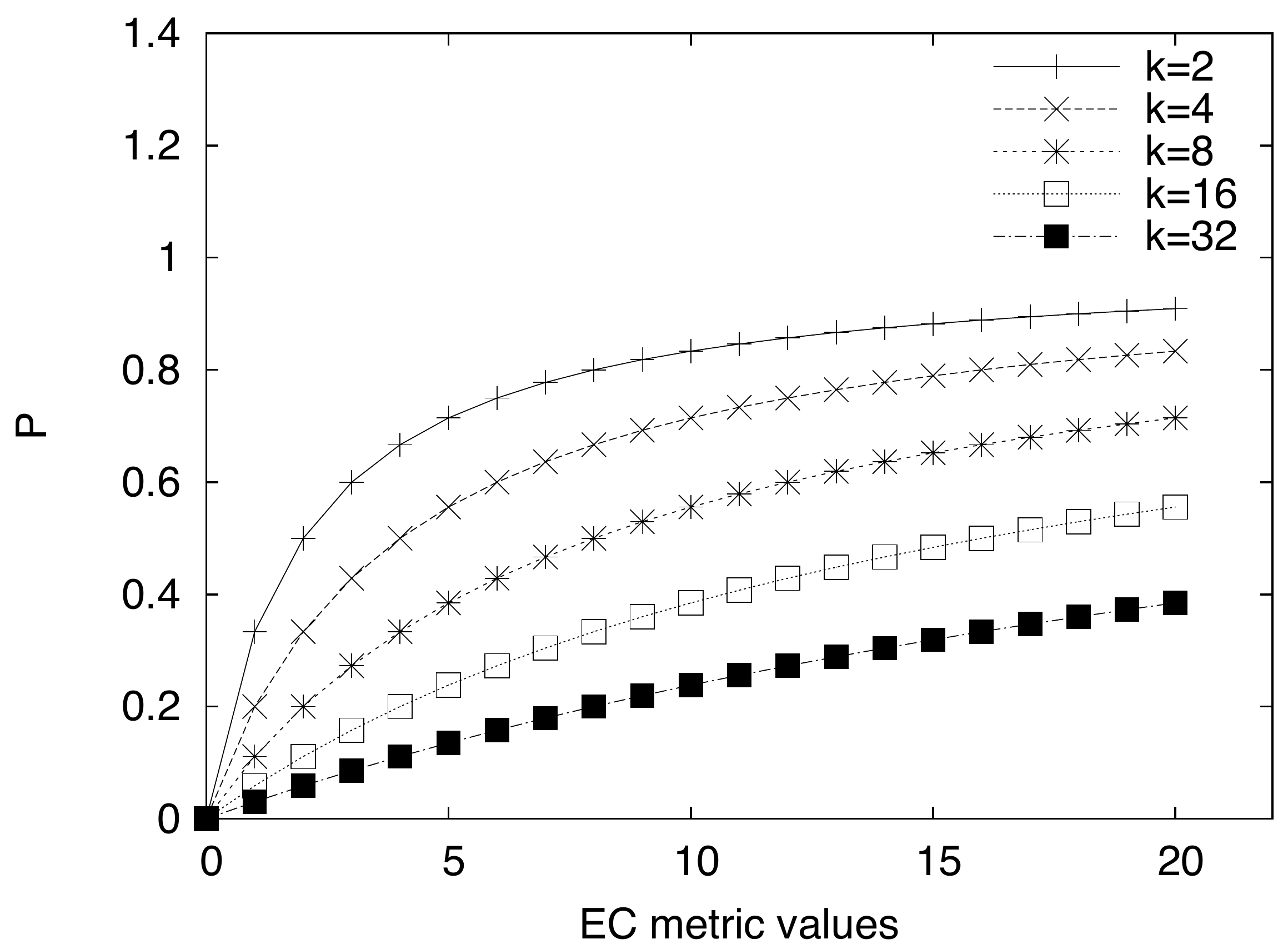}
  \caption{Relationship between $EC$ and $P$ for various $k$}
  \label{fig:P1}
\end{figure}

In both cases, the proportion of change $P$ lies between zero and one, and is calculated as:

\vspace{-0.2cm}
\begin{equation}
P = 1 - \frac{1}{\frac{metric - ideal}{k} + 1}
\end{equation}

where $metric$ denotes either $WMC$ or $EC$ as appropriate. $ideal$ is zero in
both cases. $k$ is a dampening factor for each sub-policy, a positive
constant that controls the rate of change of $P$ with respect to the difference
between the metric value and its ideal value. The higher the value of $k$, the
lower the resulting proportion of change, and hence the slower the resulting
response by the manager. This is illustrated in Fig. \ref{fig:P1}, which shows
the proportion of change resulting from various $EC$ metric values, for a range 
of $k$ values.
The overall response of the policy is to set the maintenance interval to the
mean of the values recommended by the sub-policies. In addition, if any error
has been detected during the current cycle, a maintenance operation is
invoked immediately. This is intended to improve reaction to phase changes when
the error rate increases rapidly. Without this action, errors occurring during a
long maintenance interval (set due to a currently low error rate) would not be
rectified until the end of the interval.\footnote{Note that the length of a
maintenance interval may be considerably longer than the duration of an
autonomic cycle.}

\subsection{\label{impl}Execution}

Once the planning phase is complete, the execution phase involves invoking a
maintenance operation if necessary, and setting the new value for the Chord
node's maintenance interval. The latter is achieved simply via a
{\tt{setMaintenanceInterval()}} interface added to the Chord node
implementation.

\section{\label{validation}Experiment Design}

\subsection{Overview}

The effects of the autonomic manager on Chord's performance and network usage
were measured in a sequence of experiments. Each experiment combined a
particular:

\begin{itemize}
\item workload (a temporal pattern of lookup requests),
\item membership churn (a temporal pattern of nodes joining and leaving the overlay), and
\item scheduling policy.
\end{itemize}

For each experiment, a fixed number of Chord nodes were deployed in an isolated
test-bed, each on a separate physical machine to avoid competition for CPU and
network resources on individual machines. To test repeatability of results,
each experiment was repeated three times.

\subsection{Workloads}

Each workload was specified as a temporal pattern of P2P routing requests (termed
\emph{lookup} operations). The keys being looked up were pseudo-randomly
generated with a fixed seed for each experiment.

\begin{itemize}

\item A \emph{synthetic light-weight workload} represented scenarios in which few
lookups were executed. A total of 10 lookups were issued, with 300 seconds of
inactivity between each one.

\item A \emph{synthetic heavy-weight workload} represented scenarios in which
lookups were executed at a high rate. A total of 6,000 lookups were issued, with
no delay between each one.

\item A \emph{synthetic variable-weight workload} represented scenarios involving
temporal variation in workload intensity. A total of 1,000 lookups were issued,
in batches of 100 successive lookups followed by 300 seconds of inactivity.
  
\item A \emph{file system workload} simulated the Chord workload that might
result from running a real world file system trace on a file system built above
Chord.\footnote{This research took place within the context of work using a P2P
overlay as a platform for a distributed file system.} The resulting workload
contained 15,000 lookups.

\end{itemize}

Whereas the synthetic workloads were all sequential, the file system workload
contained a mix of sequential and parallel lookups, since the lookups required
for some file system operations (such as locating all replicas of a file) can be
performed in parallel.

\subsection{Churn Patterns}

Each churn pattern modeled the behavior of a set of nodes, in terms of a
sequence of alternating on-line and off-line phases for each node.

The first two churn patterns represented uniform behavior among all nodes, one
pattern with a low churn rate and the other with a high churn rate. The durations
of the on-line and off-line phases were pseudo-randomly generated from specified
normal distributions.

Two other churn patterns represented networks exhibiting both low and high churn
rates. In one, the behavior of any given node exhibited either low or high
churn consistently throughout the experiment. The final churn pattern involved
a series of phases during which the whole network switched repeatedly between
low and high churn rates.

\begin{itemize}

\item In the \emph{low churn pattern} all nodes exhibited a sequence of
alternating on-line and off-line phases with durations drawn from the
following normal distributions:

	\begin{itemize}
	  \item on-line: $\mu=10,000s, \sigma=0s$
	  \item off-line: $\mu=160s, \sigma=20s$
    \end{itemize}
    
For each node it was pseudo-randomly decided whether it started in an on-line
or off-line phase. Since the on-line phases were chosen to be longer than the
overall experiment duration, the only variation between nodes arose from the
initial off-line phase, if present.

\item In the \emph{high churn pattern} all nodes exhibited a sequence of
alternating on-line and off-line phases with durations drawn from the
following normal distributions:

	\begin{itemize}
	  \item on-line: $\mu=200s, \sigma=40s$
	  \item off-line: $\mu=100s, \sigma=20s$
    \end{itemize}

\item In the \emph{locally varying churn pattern} 25\% of the nodes exhibited
the low churn behavior described above, and 75\% of the nodes exhibited
high churn behavior.

\item In the \emph{temporally varying churn pattern} the entire network
exhibited alternating phases of low churn and high churn behavior. The duration
of each phase was $\approx1,000$s.

\end{itemize}

\subsection{Scheduling Policies}

The following policies for maintenance scheduling were used:

\begin{itemize}

    \item A null policy, \emph{policy 0}, made no dynamic changes to the
    maintenance interval. For fair comparison, this was implemented using the
    same mechanisms as the autonomic policies. Thus the policy was invoked in
    the same way as the others, incurring the same management overheads.
    
    \item A `relaxed' autonomic policy, \emph{policy 1}, used high dampening
    factors (8 for $WMC$ and 32 for $EC$), yielding a relatively slow response
    to unsatisfactory situations.

    \item An `aggressive' autonomic policy, \emph{policy 2}, used low dampening
    factors (1 for both $WMC$ and $EC$), yielding a relatively rapid
    response to unsatisfactory situations.

\end{itemize}

In all cases the duration of the autonomic cycle was set at 2s, thus the
policies were evaluated every 2s. The initial default maintenance interval was
also set at 2s.

\subsection{\label{c:p2p_exp::ULM_spec}Evaluation Criteria}

The effectiveness of the various policies was evaluated in terms of impact on
Chord performance as perceived by the user, and on network traffic generated by
Chord. Network usage was measured simply as the mean outgoing data rate
for all nodes.

The chosen performance metric was \emph{expected lookup time}, defined as the
mean overall duration of lookup operations, under the assumption that the caller
retries repeatedly on error until a result is obtained.

Values for this metric were derived from the following measurements extracted
from the experimental logs:

\begin{itemize}

\item \emph{lookup time} $t_{lookup}$, the mean duration of individual
successful lookup operations

\item \emph{lookup error time} $t_{error}$, the mean duration of individual
failed lookup operations

\item \emph{lookup error rate} $p_{error}$, the probability that an
individual lookup operation fails

\end{itemize}

The \emph{expected lookup time} comprised the cost of the eventual successful
lookup plus the weighted sum of all possible sequences of successive failures:

\begin{equation}
	t_{lookup} + \sum\limits_{i=1} ^ \infty i \times t_{error} \times p_{error} ^ i
\end{equation}

We calculated two different versions of the \emph{expected lookup time} ($ELT$)
and \emph{network usage} ($NU$) metrics. In one version, a value was calculated
for each metric for each successive 5 minute time-window during the course of each
experiment. This allowed us to plot the metric values over time. The
disadvantage of this version was that there were some time-windows for which no
$ELT$ value could be calculated, since no successful lookup operations were
performed during the time-window.

The other version of the metrics involved calculating a single value for each
metric over the entire course of each experiment. This gave well-defined values
in all cases, at the cost of no longer allowing any insight into changes over
time.

\subsection{Experiment Platform}

The experiments were conducted on a local area test-bed consisting of 16
dedicated hosts each with a 3GHz Intel\textregistered Pentium\textregistered 4
CPU and 1GB of RAM. The hosts were connected to a dedicated switch and isolated
from the rest of the network. A separate host, the \emph{workload-executor}, ran
the workload and recorded the performance measurements needed to derive $ELT$.
Network usage measurements were recorded locally on each node, and collected
after the experiments to reduce probe effects.

This style of experiment platform was chosen to allow us to focus on obtaining
repeatable results for a realistic small-scale deployment, this being of
immediate interest in the storage research that led to this work.

\section{\label{results}Results}

\subsection{Overview}

The experiments comprised all combinations of the four workloads and four churn
patterns. Table \ref{tab:p2p_winners} shows the number of experiments in which
each policy yielded the best results, for $ELT$, $NU$, and for both
together.\footnote{Raw data from the experiments is available from
the authors on request.}

\begin{table}[h!]
\begin{center}
\begin{tabular}{|l||c|c|c|}
\hline
 & $ELT$ & $NU$ & both \\\hline
{\bf{policy 0}} (null) & 2 (\emph{2}) & 1 (\emph{0}) & 0 (\emph{0}) \\\hline
{\bf{policy 1}} (relaxed) & 8 (\emph{6}) & 3 (\emph{3}) & 0 (\emph{1}) \\\hline
{\bf{policy 2}} (aggressive) & 6 (\emph{8}) & 12 (\emph{13}) & 5 (\emph{7})
\\\hline
\end{tabular}
\caption{Number of experiments `won' by each policy, assessed using time-window
metrics and, in brackets, single-value metrics}
\label{tab:p2p_winners}
\end{center}
\end{table}

Table \ref{tab:p2p_effect_summary} shows the values of the $ELT$ and $NU$
metrics in the autonomically managed systems, normalized relative to the
unmanaged system. Thus values below one represent improvements achieved by
autonomic management. The first number given in each case is the mean of all the
time-window metric values, while the second, in italics, is the single-value
metric. Bold numbers highlight cases where significantly different values were
obtained for the different versions of the metrics.

In most cases (54 or 52 out of the 64 comparisons, for time-window and
single-value respectively), autonomic management yielded better results than the
unmanaged system, by detecting unsatisfactory situations and adapting the
maintenance interval accordingly.

The more aggressive policy, \emph{policy 2}, was the better of the two autonomic
policies. It gave an improvement in both performance and resource consumption for
75\% of combinations tested, and an improvement in one and worsening in the other
for 25\% of combinations; none of the combinations led to a worsening in both
metrics. The average \emph{expected lookup time} was 90\%
(187\%)\footnote{Figures using single-value metrics are given in brackets.} of
the average for the unmanaged system, while the resource consumption was 46\%
(39\%).

The large difference in $ELT$ values for the two versions of the metric
demonstrates that the mean of all time-window values was not representative of
overall performance. This is because the time-window version masked the two cases
in which \emph{policy 2} gave very poor performance results: a light-weight
workload with either high or temporally varying churn. However, we think that
these poor results were largely due to an artefact of our experimental design,
and the resulting methodology for calculating $ELT$. We discuss this in section
\ref{effects}, and argue that performance for real workloads would be
considerably better.

\begin{table}[h!]
\begin{center}
\begin{tabular}{|l|l||p{0.8cm}|p{0.8cm}||p{0.8cm}|p{0.8cm}|}
\cline{3-6}
\hline
\multicolumn{2}{|c||}{}&
\multicolumn{2}{c||}{\bf{policy 1}}&
\multicolumn{2}{c|}{\bf{policy 2}}\\\hline
\multicolumn{1}{|m{1.1cm}|}{\emph{workload}}&
\multicolumn{1}{m{1.2cm}||}{\emph{churn}}&
\multicolumn{1}{m{0.3cm}|}{$ELT$}&
\multicolumn{1}{m{0.3cm}||}{$NU$}&
\multicolumn{1}{m{0.3cm}|}{$ELT$}&
\multicolumn{1}{m{0.3cm}|}{$NU$}\\\hline
\multirow{8}{1.2cm}{light-weight}
	&\multirow{2}{*}{low}&0.72  &0.09 &0.70 &0.03\\\cline{3-6}
		&&\emph{0.73} &\emph{0.09} &\emph{0.70} &\emph{0.03}\\\cline{2-6}
	&\multirow{2}{*}{high} &\bf{0.81} &0.55  &\bf{0.82} &0.35 \\\cline{3-6}
		&&\bf{\emph{13.4}} &\emph{0.55} &\bf{\emph{14.0}} &\emph{0.35}\\\cline{2-6}
	&\multirow{2}{*}{local}  &0.83 &0.45 &1.21  &0.44 \\\cline{3-6}
		&& \emph{0.89} &\emph{0.45} & \emph{1.24} & \emph{0.44}\\\cline{2-6}
	&\multirow{2}{*}{temporal} &0.81  &0.29  &\bf{0.98 } &0.18 \\\cline{3-6}
		&& \emph{0.73} &\emph{0.29} &\bf{\emph{4.21}} &\emph{0.18}\\\hline
\multirow{8}{1.2cm}{heavy-weight}
	&\multirow{2}{*}{low} &0.73  &0.31  &0.70 &0.23 \\\cline{3-6}
		&& \emph{0.73} & \emph{0.22} & \emph{0.70} & \emph{0.16}\\\cline{2-6}
	&\multirow{2}{*}{high} &0.60 &1.33 &0.69 &0.98 \\\cline{3-6}
		&& \emph{0.57} & \emph{1.29} & \emph{0.61} & \emph{0.95}\\\cline{2-6}
	&\multirow{2}{*}{local}&\bf{0.08 } &\bf{1.27 } &\bf{0.18 } &\bf{1.08 }\\\cline{3-6}
		&&\bf{\emph{0.36}} &\bf{\emph{0.56}} &\bf{ \emph{0.60}} &\bf{\emph{0.67}}\\\cline{2-6}
	&\multirow{2}{*}{temporal}&0.56 &0.54 &0.67 &0.40 \\\cline{3-6}
		&& \emph{0.68} & \emph{0.39} & \emph{0.79} & \emph{0.34}\\\hline
\multirow{8}{1.2cm}{variable}
	&\multirow{2}{*}{low} &0.71 &0.11 &0.70  &0.05 \\\cline{3-6}
		&& \emph{0.71} & \emph{0.10} & \emph{0.70} & \emph{0.05}\\\cline{2-6}
	&\multirow{2}{*}{high} &0.36 &1.20 &0.36 &0.78 \\\cline{3-6}
		&& \emph{0.39} & \emph{1.24} & \emph{0.39} & \emph{0.81}\\\cline{2-6}
	&\multirow{2}{*}{local}&\bf{3.24 } &0.42 &\bf{2.95 } &0.42 \\\cline{3-6}
		&&\bf{\emph{1.52}} & \emph{0.47} &\bf{ \emph{1.75}} & \emph{0.47}\\\cline{2-6}
	& \multirow{2}{*}{temporal}&1.56 &0.26 &0.97 &0.24 \\\cline{3-6}
		&&\emph{1.52} &\emph{0.29} &\emph{0.94} & \emph{0.24}\\\hline
\multirow{8}{1.2cm}{file-system}
	&\multirow{2}{*}{low} &0.80 &0.34 &0.79 &0.29 \\\cline{3-6}
		&& \emph{0.80} &\emph{0.26} & \emph{0.79} & \emph{0.22}\\\cline{2-6}
	&\multirow{2}{*}{high} &\bf{5.14 } &\bf{0.47 } &1.09  &0.52 \\\cline{3-6}
		&& \bf{\emph{2.42}} &\bf{\emph{1.70}} & \emph{1.00} & \emph{0.47}\\\cline{2-6}
	&\multirow{2}{*}{local} &0.60 &0.45 &0.59 &0.88 \\\cline{3-6}
		&& \emph{0.70} & \emph{0.28} & \emph{0.57} & \emph{0.46}\\\cline{2-6}
	&\multirow{2}{*}{temporal} &0.86 &0.59  &0.93  &0.41 \\\cline{3-6}
		&&\emph{0.86} &\emph{0.51} & \emph{0.91} & \emph{0.40}\\\hline\hline
\multirow{4}{1.2cm}{summary}
	&\multirow{2}{*}{\bf{mean}} &1.15 &0.54 &0.90 &0.46 \\\cline{3-6}
		&& \emph{1.69} & \emph{0.54} & \emph{1.87} & \emph{0.39}\\\cline{2-6}
	&\multirow{2}{*}{\bf{median}} &0.77 &0.45 &0.75 &0.41 \\\cline{3-6}
		&& \emph{0.73} & \emph{0.42} & \emph{0.79} & \emph{0.38}\\\hline
\end{tabular}
\caption{Normalized performance and network usage metrics (single-value
metrics shown in italics, significant differences between single-value and
time-window metrics shown in bold)}
\label{tab:p2p_effect_summary}
\end{center}
\end{table}
\newpage
\subsection{Autonomic Manager Behavior}

\begin{figure*}[t]
  \begin{center}
    \subfloat[Low churn\label{fig:WL2_NB1_averaged_fixFinger_interval}] 
    {\includegraphics[width=8cm]{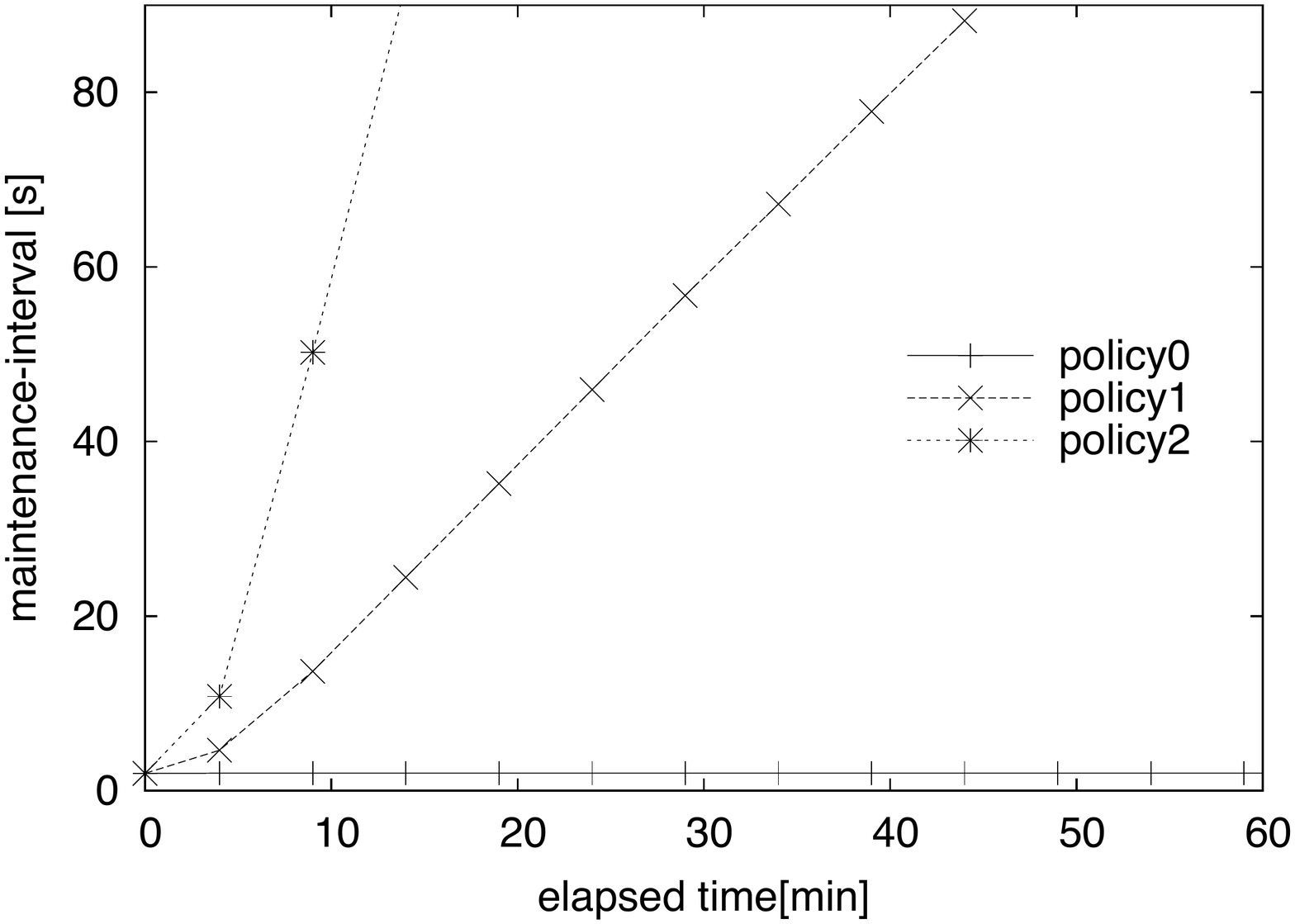}}
    \hspace{5mm}
    \subfloat[High churn\label{fig:WL2_NB2_averaged_fixFinger_interval}]
    {\includegraphics[width=8cm]{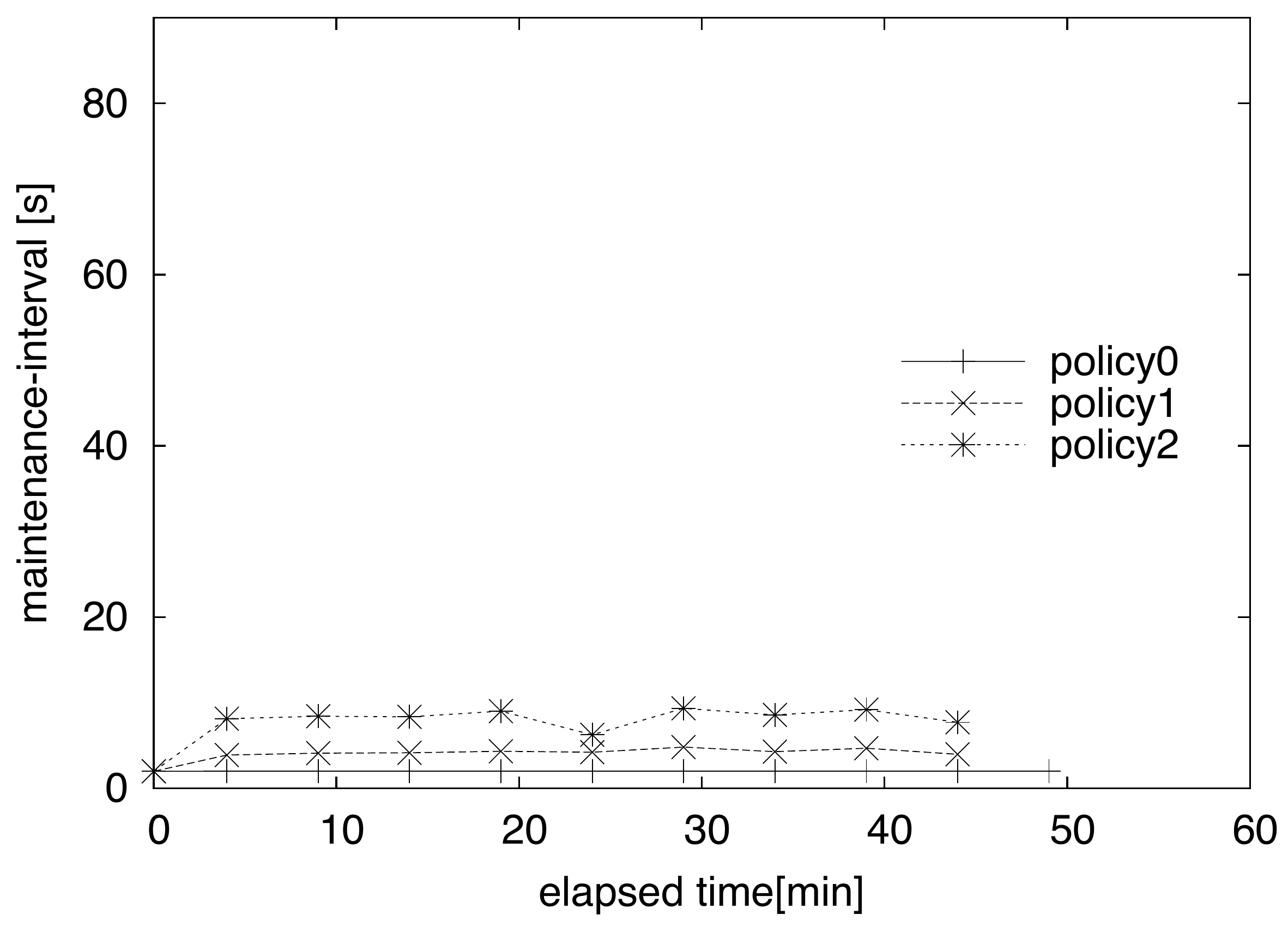}}
    \vspace{5mm}
    \subfloat[Locally varying churn\label{fig:WL2_NB3_averaged_fixFinger_interval}] 
    {\includegraphics[width=8cm]{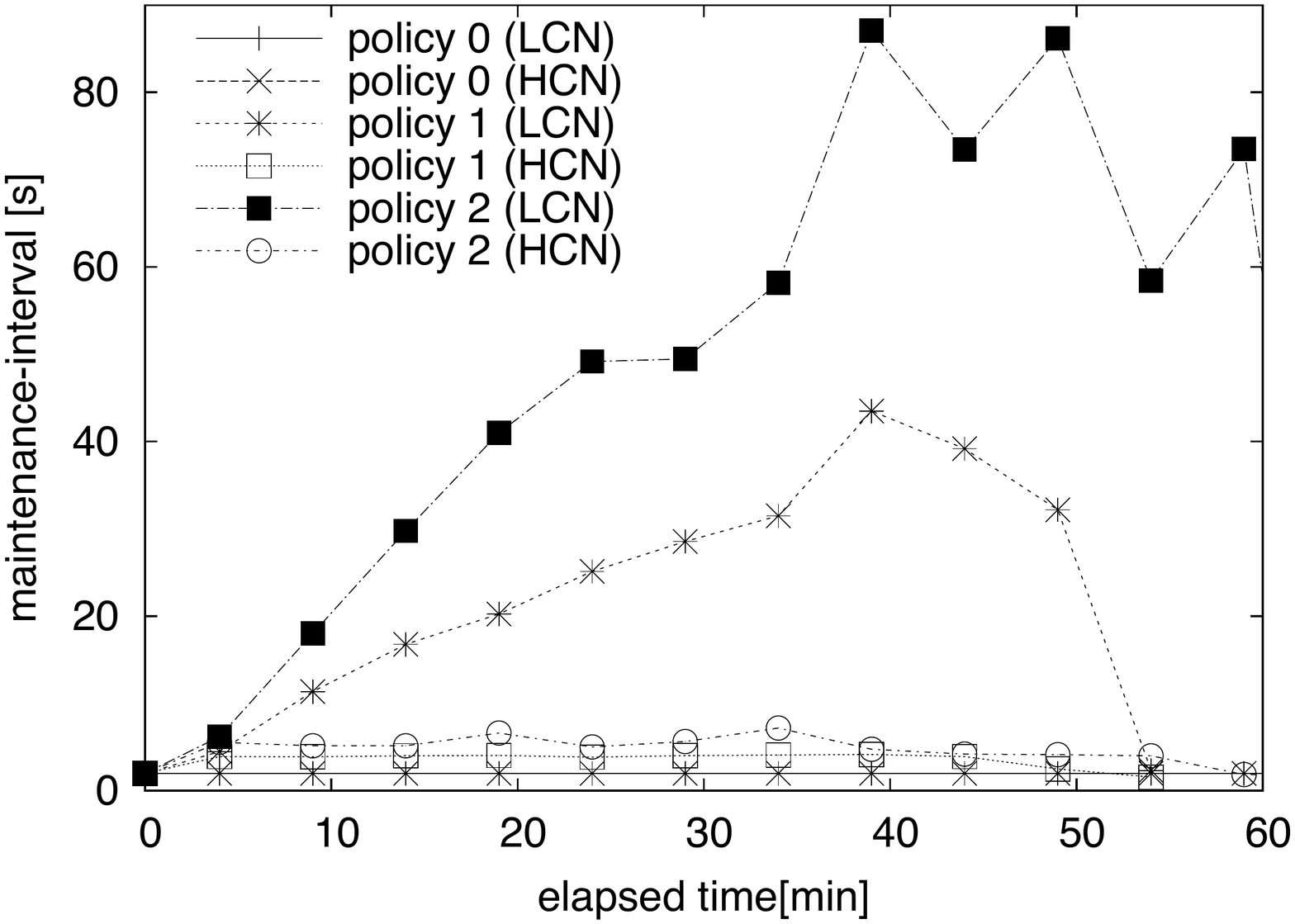}}
    \hspace{5mm}
    \subfloat[Temporally varying churn\label{fig:WL2_NB4_averaged_fixFinger_interval}] 
    {\includegraphics[width=8cm]{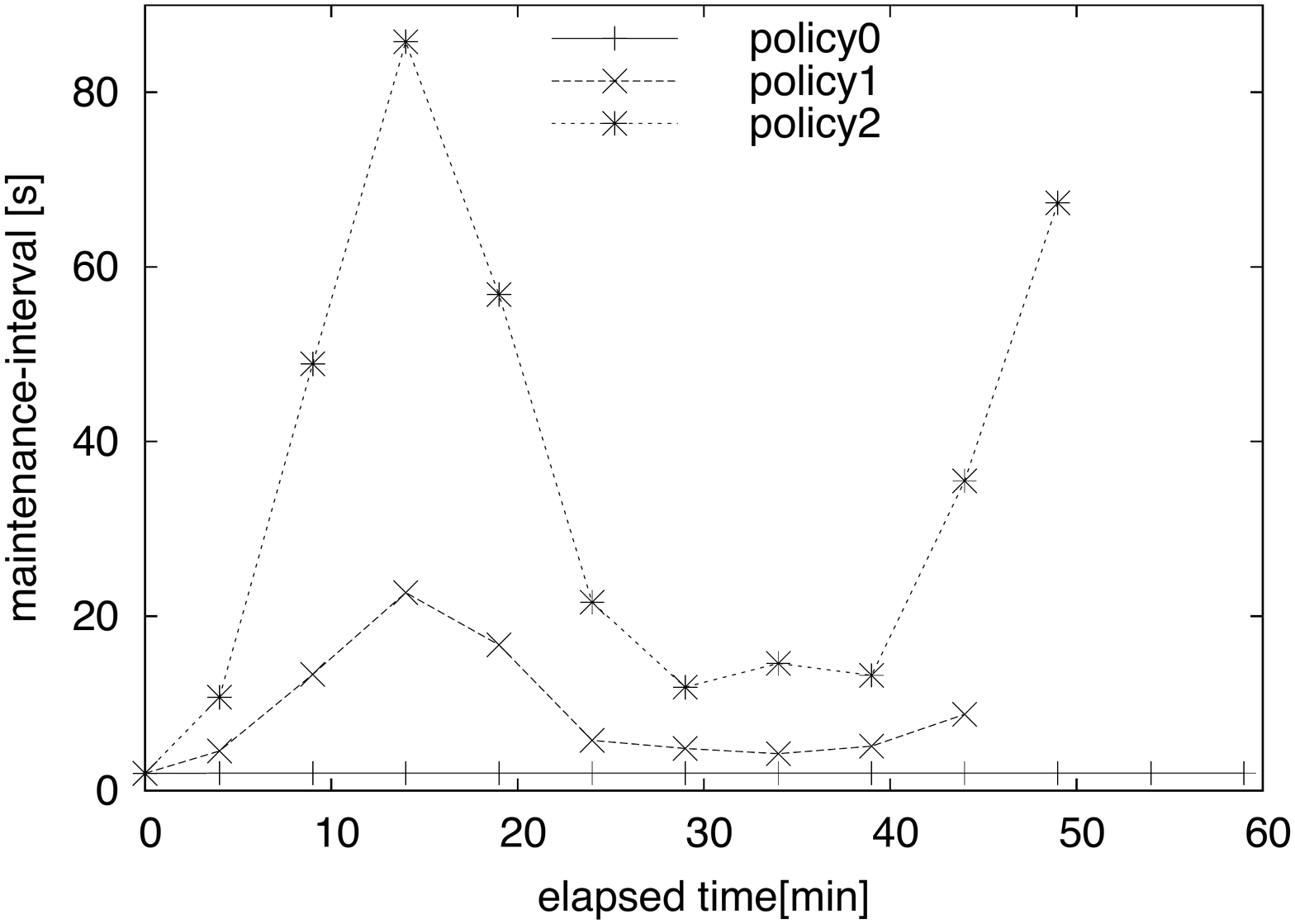}}
  \end{center}
\vspace{-1mm}
\caption{Interval progressions for heavy-weight workload and various churn
patterns}
\end{figure*}

We now examine in more detail the effects of \emph{policy 2} in a sample group of
experiments, those using a heavy-weight workload. Table
\ref{tab:p2p_effect_summary} shows that in the experiments with low, high and
temporally varying churn, \emph{policy 2} yielded greater performance and reduced
resource consumption. For locally varying churn the effect varies depending on
which of the metric versions is considered. To illustrate the policy actions
resulting in these effects, we plot the progression of maintenance intervals over
time. Figs
\ref{fig:WL2_NB1_averaged_fixFinger_interval}-\ref{fig:WL2_NB4_averaged_fixFinger_interval}
show the progressions of the maintenance intervals over the courses of the
experiments, for the four churn patterns. Each point plotted is the mean of the
corresponding figures for three repeated runs. In the experiment with locally
varying churn, the progressions are plotted separately for low-churn nodes (LCN)
and high-churn nodes (HCN).

Fig. \ref{fig:WL2_NB1_averaged_fixFinger_interval} shows that for low churn, the
autonomic policies detected an unsatisfactory situation with respect to
\emph{network usage}, and reacted by steadily increasing the maintenance
interval. This decreased the amount of work each node spent (unnecessarily) maintaining its
peer-set, and thus reduced the amount of data sent to the network in comparison
with unmanaged nodes. Additionally, a reduction in the work spent on maintenance
operations left more computational capacity for dealing with lookup operations.
This reduced the \emph{expected lookup time}. The progressions of the
\emph{network usage} and \emph{expected lookup time} metrics are described later in this
section. As expected, \emph{policy 2} reacted more aggressively, increasing the
maintenance interval at a higher rate than \emph{policy 1}.

Fig. \ref{fig:WL2_NB2_averaged_fixFinger_interval} shows that for high churn,
the autonomic policies held the intervals fairly constant, though at higher
values than for unmanaged nodes. Referring to Table
\ref{tab:p2p_effect_summary}, this yielded roughly the same \emph{network usage}
as for unmanaged nodes, and a significant improvement in \emph{expected lookup time}.
This improvement in performance may appear counter-intuitive, given the
reduction in overall maintenance effort, particularly since examination of the
experimental logs shows that error rates were significantly \emph{lower} for the
autonomic policies than for unmanaged nodes. The explanation is that each value
plotted is derived by averaging individual maintenance interval values over the
entire network, and over a five minute aggregation time window. This masks the
fact that there was considerable variation in controlled interval values within
each time window. Whenever the manager of a given node detected errors in its
peer-set, it immediately decreased the maintenance interval, giving a period of
high maintenance activity. Once the errors were corrected, the manager increased
the interval again until the next error. Thus errors were corrected more
rapidly than in an unmanaged system, despite the overall average interval being
higher. Again, \emph{policy 2} reacted more aggressively than \emph{policy 1},
and kept the maintenance interval at higher levels.

Fig. \ref{fig:WL2_NB3_averaged_fixFinger_interval} shows the resulting intervals
for locally varying churn, where some nodes (75\%) exhibited high churn, and the
rest, low churn. There are two features of interest: the apparent phase change
after about 40 minutes, and the fact that the autonomic managers behaved
markedly differently on the low churn and high churn nodes.

We have no simple explanation for the phase change, other than to hypothesize
that the particular churn patterns in use caused some threshold in the error
rate to be exceeded, triggering rapid decreases in the intervals.

The differences in behavior between low and high churn nodes appear anomalous,
since all nodes experience roughly the same environment in terms of the aggregate
behavior of their peers (assuming that the low churn nodes are uniformly
distributed throughout the network). The explanation is that a node's maintenance
interval was reset to the default value every time it restarted, in order to
simulate the arrival of new nodes in a network. Thus each manager on a high churn
node did react to the environment in the same way as the low churn nodes, by
steadily increasing the maintenance interval, but since the interval was
regularly reset, the average value was held fairly constant and close to the
default value.

We could have chosen not to reset the interval on restart, to simulate nodes
failing and recovering rather than having nodes permanently leaving and new
nodes joining in their place. In that case we would expect to see little
difference in the behavior of low and high churn nodes.

Fig. \ref{fig:WL2_NB4_averaged_fixFinger_interval} shows the autonomic policy
behavior for temporally varying churn, in which the entire network alternated
between low and high churn, in phases lasting about 17 minutes. During the
initial low churn phase the managers responded as expected, in the same way as in
the low churn experiment. When the network moved into high churn they reacted by
decreasing the maintenance intervals. The more aggressive behavior of
\emph{policy 2} can be seen clearly. Table \ref{tab:p2p_effect_summary} shows
that this gave slightly better results for \emph{network usage} than
\emph{policy 1}, but slightly worse for \emph{expected lookup time}. Both policies obtained
significantly better results than the unmanaged network.

\subsection{\label{effects}Autonomic Manager Effects}
\begin{figure*}[t]
  \begin{center}
    \subfloat[Expected lookup time progressions with heavy-weight
    workload\newline and low
    churn\label{fig:WL2_NB1_averaged_expected_lookup_time}]
    {\includegraphics[width=8.5cm]{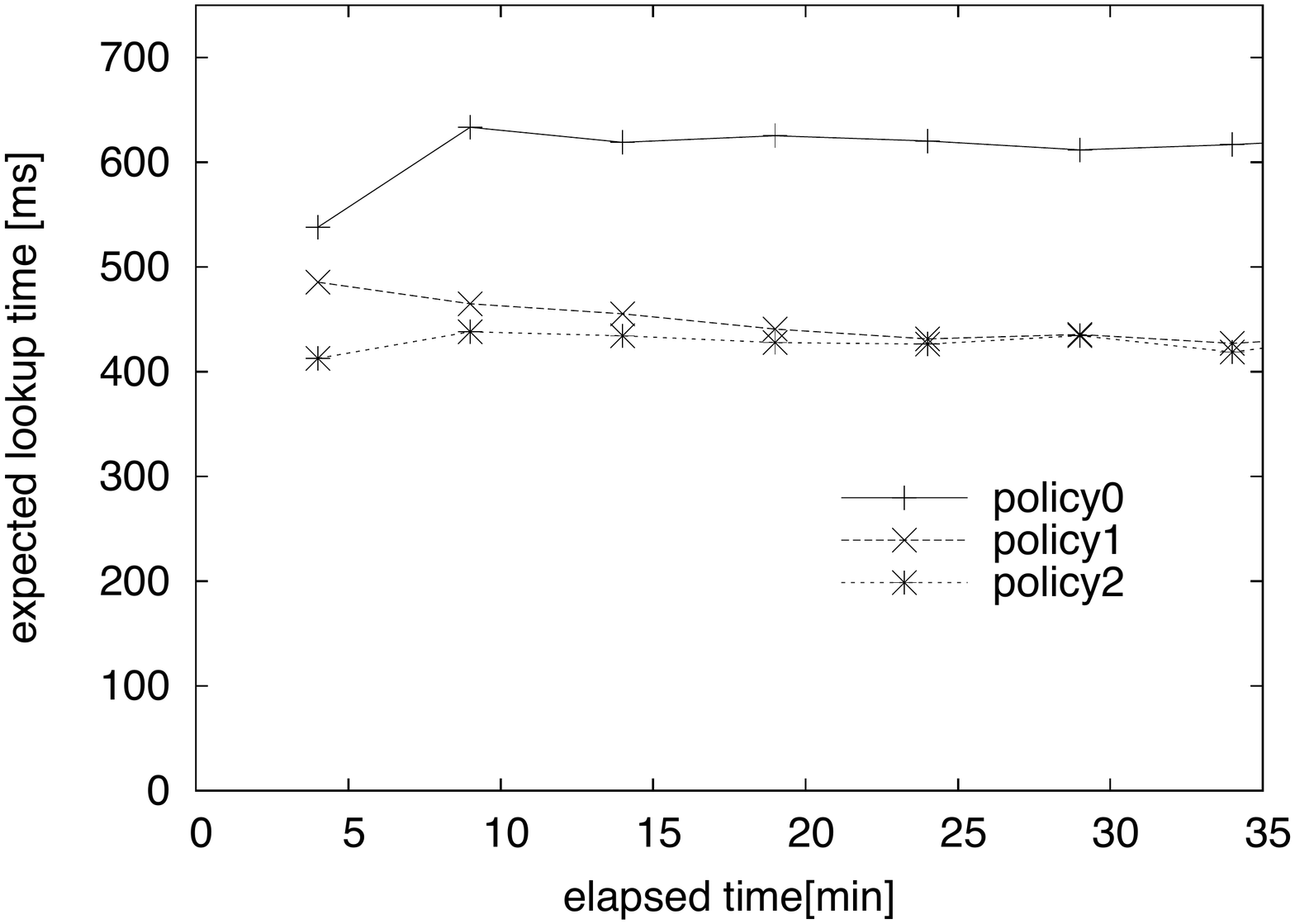}}
    \hspace{5mm}
    \subfloat[Network usage progressions with heavy-weight workload \newline and
    low churn\label{fig:WL2_NB1_averaged_network_resource_usage_rate}]
    {\includegraphics[width=8.5cm]{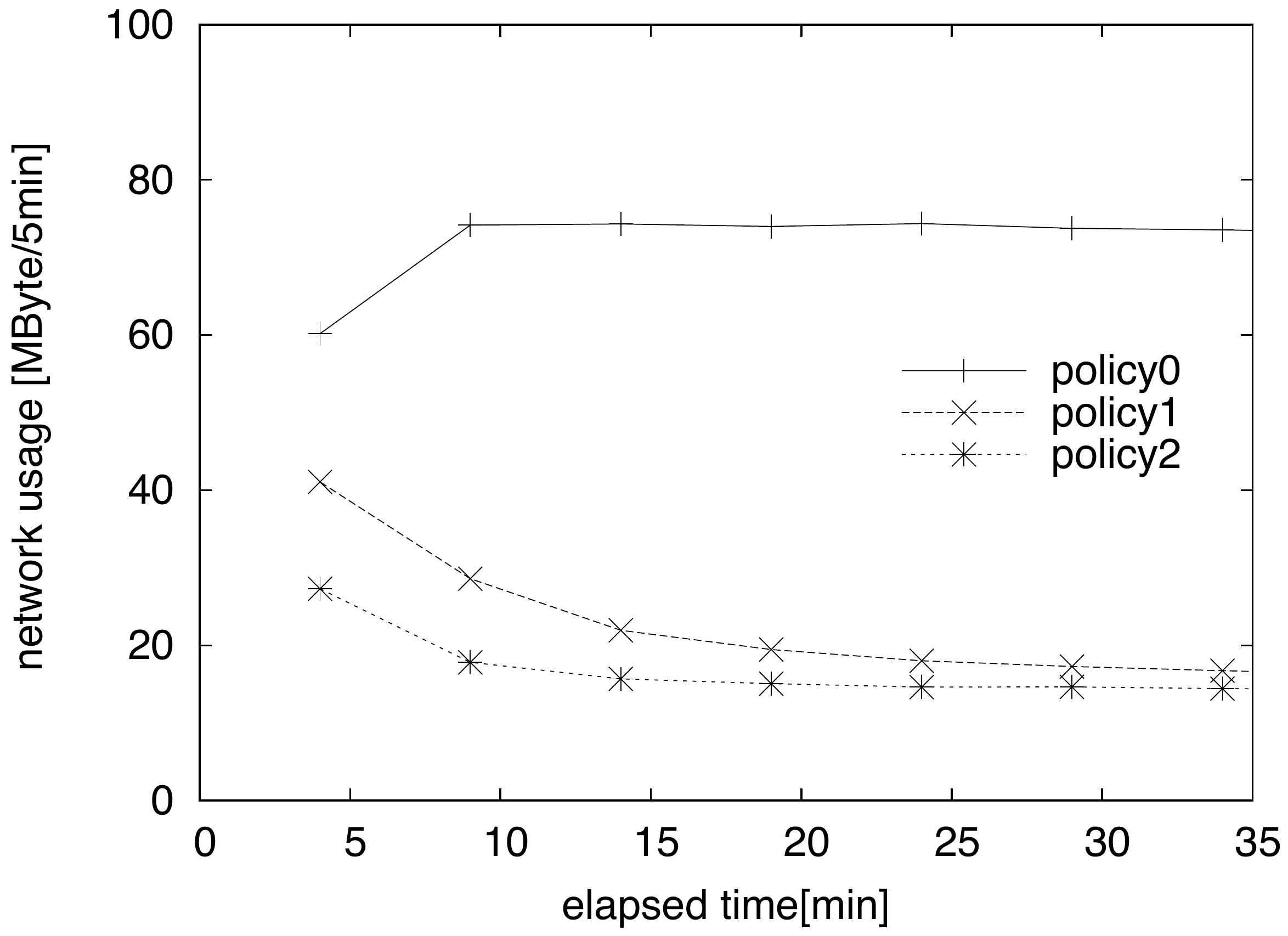}}
  \end{center}
\caption{High-level metric progressions for heavy-weight workload and low
churn}
\label{fig:metric_progression}
\end{figure*}
Fig. \ref{fig:metric_progression} shows how the time-window versions of the
high-level metrics \emph{network usage} and \emph{expected lookup time}
progressed over the course of the experiment with low churn and heavy-weight
workload. This demonstrates the concrete effects of the autonomic manager
behavior described in Fig. \ref{fig:WL2_NB1_averaged_fixFinger_interval}. For
this experiment, the two versions of the metrics give similar values, thus it is
reasonable to assume that the plotted values do correspond to what would be
experienced by a user.

Fig. \ref{fig:WL2_NB1_averaged_expected_lookup_time} shows that, under low churn,
the \emph{expected lookup time} for autonomically managed nodes stabilized,
within a few minutes, at about 70\% of the figure for unmanaged nodes. Similarly,
in Fig. \ref{fig:WL2_NB1_averaged_network_resource_usage_rate} it can be seen
that the \emph{network usage} for autonomically managed nodes stabilized early
in the experiment, this time at about 30\% of the figure for unmanaged nodes. Further
analysis showed that the \emph{network usage} with managed nodes was mainly due
to the execution of workload lookups.

We now consider the differences between the values obtained for the single-value
and time-window versions of the high-level metrics. The overall mean normalized
performance for \emph{policy 2}, as measured by the single-value version of the
$ELT$ metric, was 1.87. The median value of the metric was 0.79; the mean was
skewed by particularly poor results for light-weight workload under both high and
temporally-varying churn.

These results are obscured by the time-window version of $ELT$, hence the
introduction of the single-value version. This is because, for the light-weight
workload, there were some time-windows during which no lookup operations
completed successfully, and thus an $ELT$ value could not be calculated
as described in section \ref{c:p2p_exp::ULM_spec}. As a result, no value from
that time-window was fed into the average, leading to an apparently much better
value for the time-window average overall. We therefore feel that the
single-value metrics give a fairer means of comparison between policies.

Nonetheless, we think that the poor single-value metric values for light-weight
workloads are artificially high, due to a poor (in retrospect) experimental
design decision. During the execution of the experimental workloads, any lookup
operation resulting in an error was logged but not retried.

One consequence of this design was that for a light-weight workload there was a
high probability that significant network topology change had occurred between
any given successive pair of lookup operations, and thus a high probability that
any given lookup operation would yield an error. Even though each error
triggered an immediate maintenance operation, the following lookup operation,
occurring a significant time later, would not receive any benefit from that
maintenance, due to network topology change during the intervening period. The
resulting high error rate leads directly to a high \emph{expected lookup time}
metric value.

In practice, with a client that immediately retries each failed lookup
operation, we expect the error rate for light-weight workloads to be much lower.
This is because retried lookups will often succeed before the next network
topology change.

It would have been better to have performed such retries as part of the
experimental workload execution. This would have enabled us to simply measure
\emph{expected lookup time} rather than having to derive it from the error rate,
and it seems likely that the resulting performance metric values would have been
significantly better than the derived values presented here.

\begin{figure}[h]
  \centering
    \includegraphics[width=8cm]{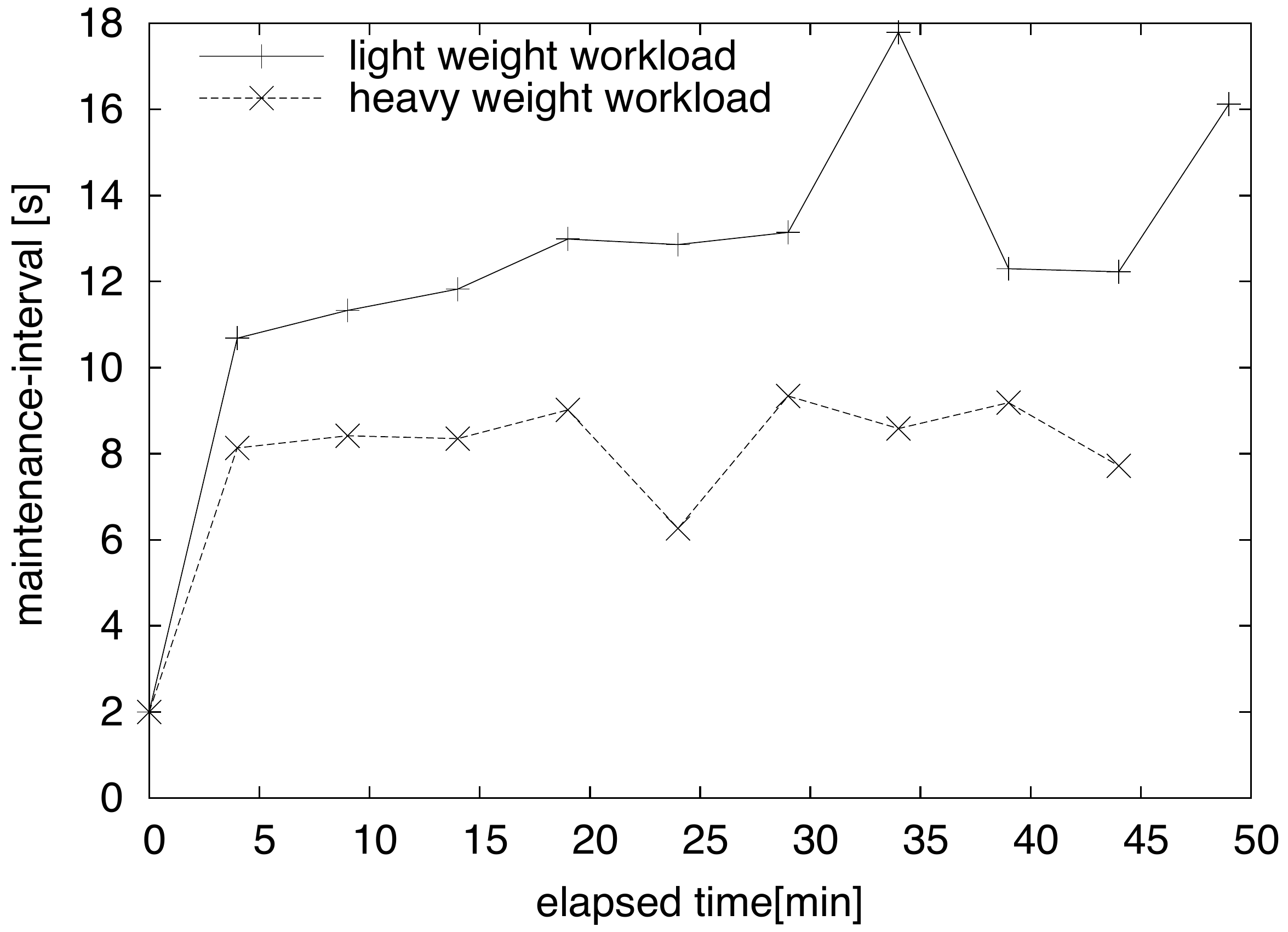}
  \caption{Interval progressions for high churn with heavy-weight and
  light-weight workloads (policy 2)}
  \label{WL1_vs_WL2_NB2_Pol2_avg_fixFinger_interval}
\end{figure}

The behavior of \emph{policy 2} for high churn and light-weight workload is
illustrated in Fig. \ref{WL1_vs_WL2_NB2_Pol2_avg_fixFinger_interval}, and
contrasted with the behavior for heavy-weight workload. 

With the light-weight workload, the autonomic manager perceives a
lower error rate due to the lower frequency of lookup operations, and thus
increases the maintenance intervals relative to those resulting from a
heavy-weight workload. It is the low absolute error rate
that governs the manager's behavior in this situation, even though the ratio of
errors to successful lookups is high. This behavior seems appropriate, since the
lighter the workload, the less effort that we wish the maintenance activity
to expend.

\subsection{Repeatability}

Each experiment was executed three times, and the results averaged to produce the
observations reported. In order to assess repeatability, the variation of the
$ELT$ values between experimental runs was investigated. For each set of three
values, corresponding to the values calculated for a particular time window in
each of the runs, a \emph{similarity metric} was calculated. This process was
performed for every experiment.

The chosen similarity metric was \emph{normalized standard deviation} ($NSD$),
defined as the standard deviation of the set of three values divided by its
mean. Thus perfect repeatability would yield zero for every $NSD$ value.

Figure \ref{fig:p2p_similarity} shows the cumulative frequency distribution of
all $NSD$ values. The median of the $NSD$ values was about 0.2; we conclude
that the observed experimental behavior was acceptably repeatable.

\begin{figure}[h]
  \centering
    \includegraphics[width=8cm]{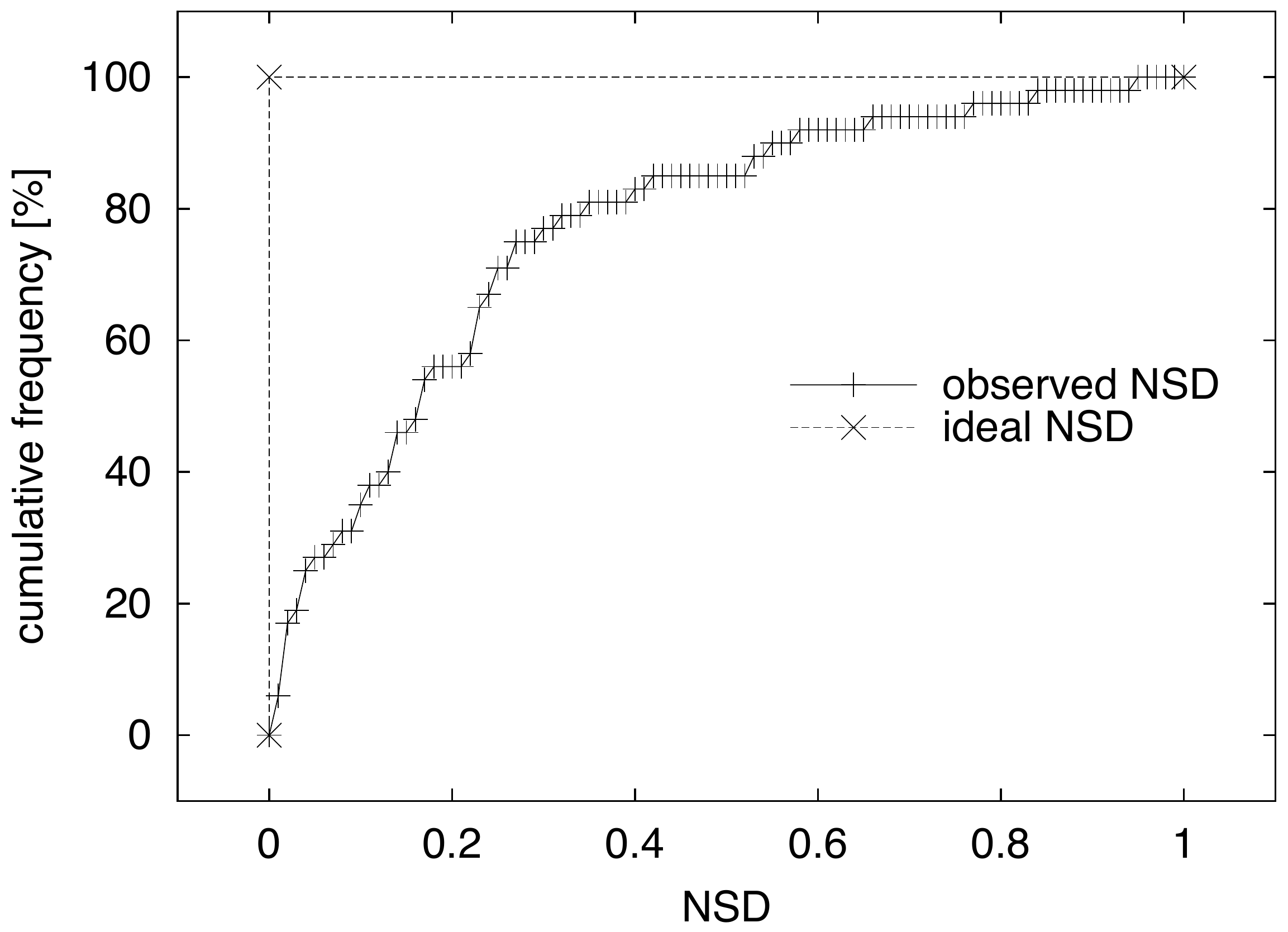}
  \caption{Cumulative frequencies for all similarity metric values}
  \label{fig:p2p_similarity}
\end{figure}

To illustrate the degree of variability represented by particular $NSD$ values,
Fig. \ref{fig:p2p_similarity} shows the $ELT$ progressions for the individual
runs of two particular experiments, with overall average $NSD$ values of
0.23 and 0.01.

\begin{figure*}[t]
  \begin{center}
    \subfloat[Average $NSD$ value of 0.23 (variable-weight workload,
  high churn, policy 2)\label{fig:p2p_similarity1}]
    {\includegraphics[width=8cm]{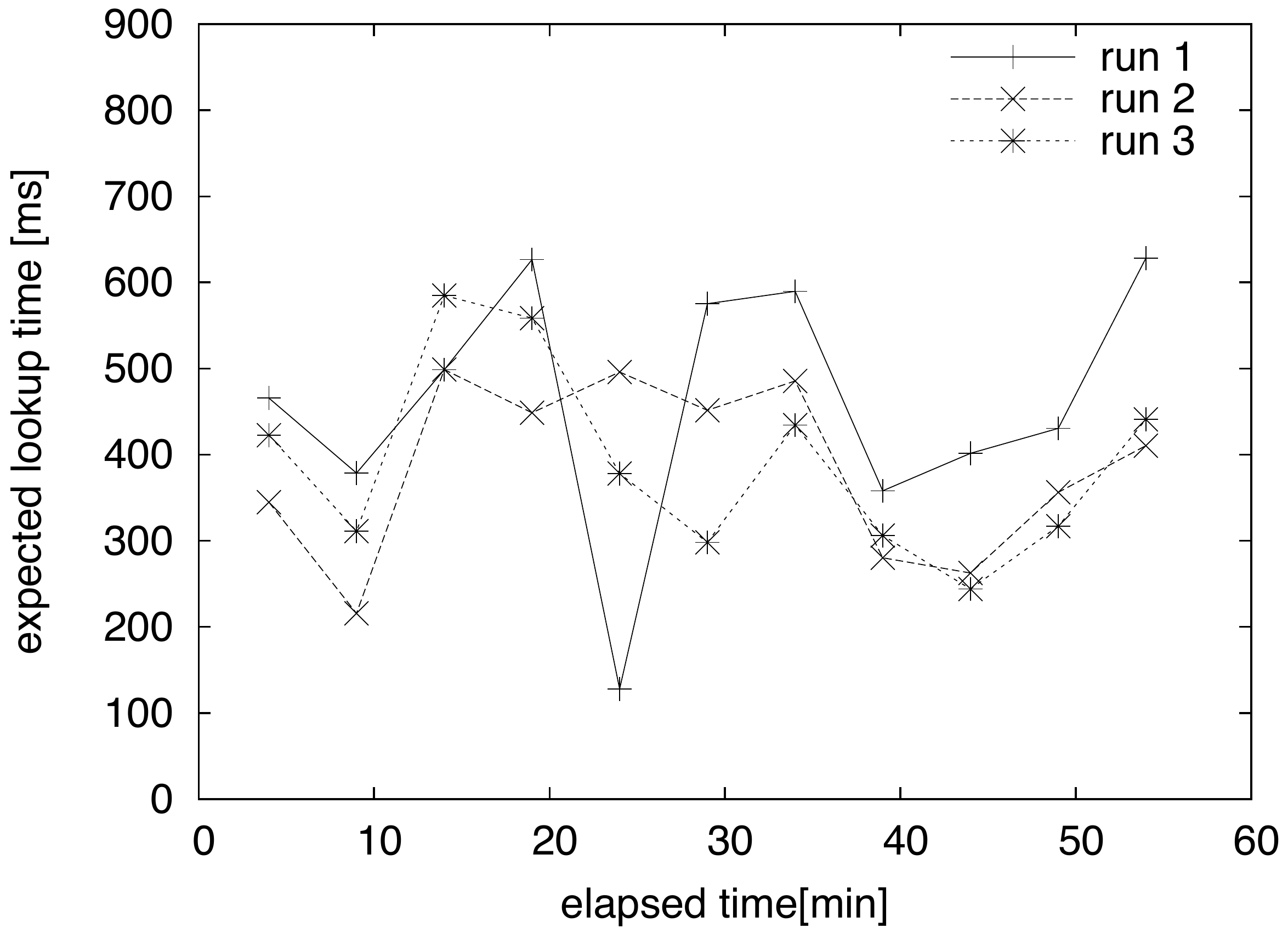}}
    \hspace{5mm}
    \subfloat[Average $NSD$ value of 0.01 (heavy-weight workload,
  low churn, policy 2)\label{fig:p2p_similarity2}]
    {\includegraphics[width=8cm]{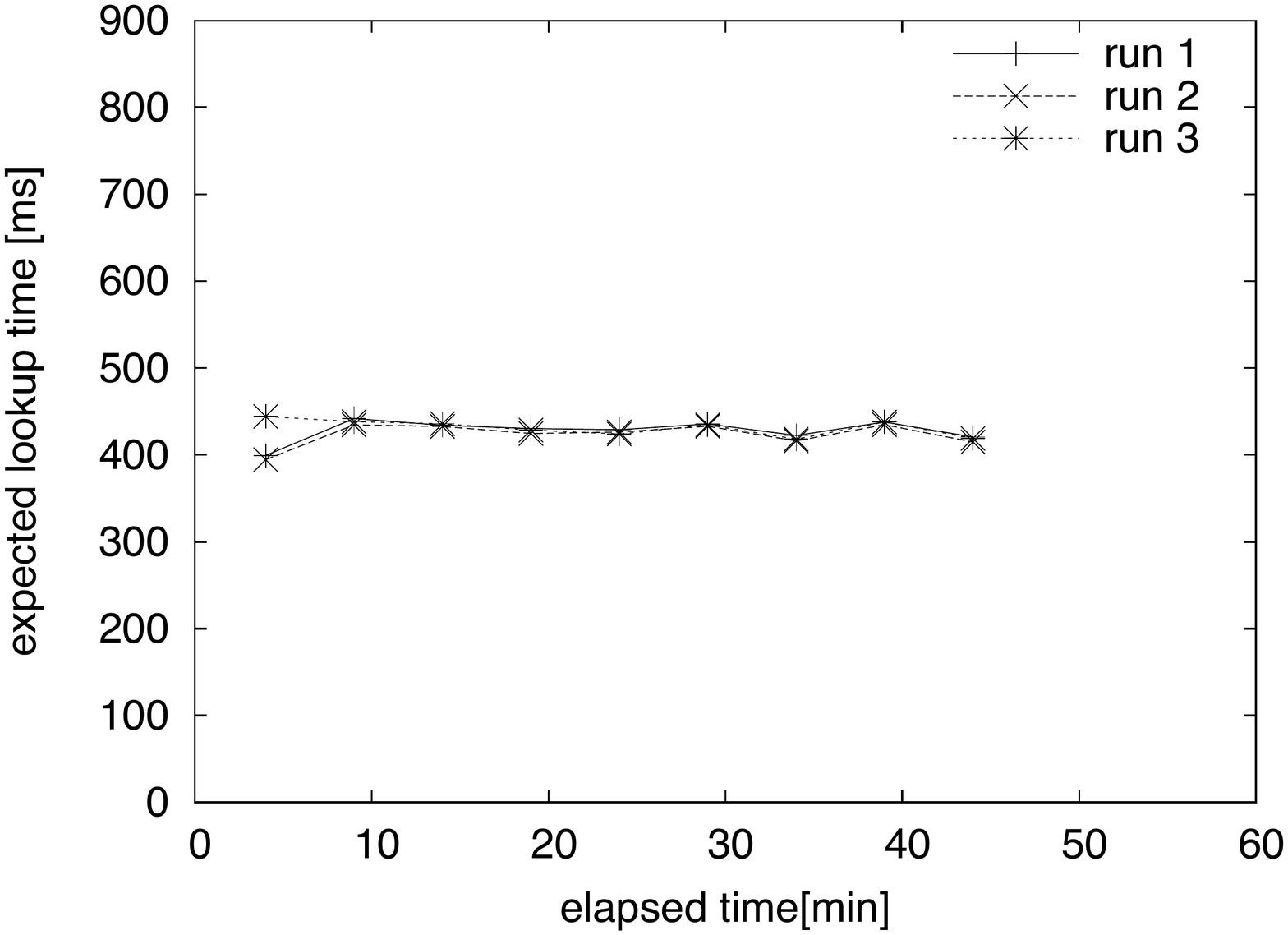}}
  \end{center}
\caption{Expected lookup time progressions for individual runs of selected experiments}
\label{fig:p2p_similarity}
\end{figure*}

\section{\label{conclusions}Conclusions and Future Work}

We have demonstrated that autonomic management of maintenance scheduling in Chord
can achieve significant improvement in performance and resource consumption in
many situations. Under changing conditions, management can adapt scheduling to
suit prevailing conditions. Under static conditions, it can converge to a better
scheduling than is likely to be configured for an unmanaged system.

We have argued that the cases in which autonomic management performed
significantly worse than an unmanaged system may be at least partially explained
by our experimental decision not to retry lookup operations on failure.
Nonetheless, there are several avenues for possible further development of our
management approach. Some involve refinements to Chord itself, while others
involve different approaches to autonomic management. Our autonomic management
approach could also be applied straightforwardly to other P2P overlay networks
that perform periodic maintenance operations.

\subsection{Chord Adaptations}

Chord could be adapted in several ways to allow autonomic management of the
structure of a node's peer-set, in addition to the scheduling of its maintenance.
The aim would be to improve lookup performance---which might be achieved by
reducing the probability of an error on a given lookup, or by reducing the number
of hops required for a given lookup.

The probability of error could be reduced by adjusting the lookup algorithm so
that it can use other fingers if the closest preceding finger to the target is
invalid. Another possibility is to cache the successor list of each finger, for
use during lookup if the finger is discovered to be invalid. The average number
of hops required could be reduced by enlarging the finger table, thus providing
fingers that are closer to the target. These options could be combined by
adding autonomic management hooks for:

\begin{itemize}
  \item the size of the finger table
  \item the number of successors to be cached for each finger 
\end{itemize}
	
As with our maintenance interval control scheme, the autonomic manager
could adjust these dynamically, thereby managing an additional tradeoff between
performance and maintenance costs. In the limit, the finger table could be
expanded to encompass the entire network. \cite{1-Hop} argues that this is
feasible, although it does not propose autonomic management of finger table size.

Another Chord aspect that could be dynamically managed is the length of each
node's successor list, as suggested in
\cite{onTheStabilityOfChordBinzenhoefer2004}. The longer the list, the greater
the number of near-simultaneous successor failures that a node can recover from,
but the greater the maintenance overhead. The autonomic manager could adjust this
based on an assessment of the recent churn level, or simply based on how many
elements of the list have been recently used for ring repair.

\subsection{Autonomic Management}

We chose to structure the autonomic management policy using distinct
sub-policies, each of which had a separate sub-goal and considered only a subset
of the monitored information available. Instead, a single policy taking a
holistic view of all monitored information could be used. This would be more
flexible but would probably also be more complex.

Our management scheme operates entirely locally; the autonomic managers work
completely independently, considering only local monitoring data, and make no
attempt to coordinate their actions. Both of these aspects could be addressed:
monitoring data could be disseminated within the network to allow managers to
take a broader view of the network state, and managers could communicate in an
effort to harmonize their actions. For example, with dynamic control of successor
list length as mentioned previously, it would be useful for a manager to know
about topology repair operations on other nodes, to allow it to better assess
the current stability of the overall network.

Our autonomic manager does not maintain long-term state. This could be added,
allowing the manager to learn from experience. The ability could be
introduced in the form of a `meta feedback loop' using a management hierarchy,
where the main manager was itself managed by a higher-level meta-manager. The
meta-manager could monitor user-centric metrics such as \emph{expected lookup
time} and \emph{network usage}, and tune the parameters of the main manager
accordingly. Such parameters could include the relative weights placed on the
sub-policies, and the dampening factors used in those sub-policies.

A different learning approach would be for the manager to periodically store a
record of the current conditions, its response to those conditions, and the
resulting effects. Given some pattern matching mechanism, it could then
periodically retrieve historical records for previous conditions similar to
those currently prevailing, and take into account the success or otherwise of
its past actions in deciding how to act in the current situation.

Finally, one other possible tactic would be for the manager to try making small
speculative adjustments and monitor their effects for a short time. This
might enable a gradient descent approach---the manager would enact the
adjustments that led in the best `direction' in terms of user-centric metrics,
and then repeat the process.

\bibliographystyle{IEEEtran}
\bibliography{IEEEabrv,asa}

\end{document}